\documentclass[aps,prl,twocolumn,10pt,superscriptaddress,nofootinbib]{revtex4-2}

\usepackage{amsmath,amssymb,amsthm,mathtools,bm,bbm}
\usepackage{graphicx}
\usepackage[colorlinks=true,linkcolor=blue,citecolor=blue,urlcolor=blue]{hyperref}
\usepackage{physics}

\newcommand{\Df}{D_f}

\begin{document}

\title{Universal Thermodynamibc Uncertainty Relation for Quantum $f-$Divergences}

\author{Domingos S. P. Salazar}
\affiliation{Unidade de Educa\c c\~ao a Dist\^ancia e Tecnologia,
Universidade Federal Rural de Pernambuco,
52171-900 Recife, Pernambuco, Brazil}
\date{\today}

\begin{abstract}
We show that any Petz $f$-divergence (where $f$ is operator convex) between quantum states admits a universal $\chi^2$–mixture representation: the distinguishability of $\rho$ from $\sigma$ is obtained as a positive superposition of quadratic contrasts $\chi^2_\lambda$, with nonnegative weights $w_f(\lambda)$ determined explicitly from the Stieltjes representation of the generator $f$. This identifies $\chi^2_\lambda$ as atomic building blocks for quantum $f$-divergences and yields closed-form $w_f$ for canonical choices (relative entropy/KL, Hellinger/Bures, Rényi). By mapping $\chi^2_\lambda$ into a classical Pearson $\chi^2$, we leverage Chapman-Robbins variational representation and obtain a tight and universal quantum thermodynamic uncertainty relation: any $f$-divergence is lower bounded by a function of the statistics of quantum observables (mean and variance), reproducing previous and novel results in quantum thermodynamics as applications.
\end{abstract}

\maketitle
\textbf{Introduction.—}
Thermodynamic uncertainty relations (TURs) have proliferated \cite{Barato2015,Gingrich2016,Horowitz2020Rev,VoV2020,Erker2017,MacIeszczak2018,Carollo2019,Goold2014b,Timpanaro2019B,Hasegawa2019,Batalhao2014}, more recently in the quantum realm \cite{LandiReview2021,Hanggi2015,Liu2019,VanVu2023,Miller2021,Pires2021,Guarnieri2019,Hasegawa2019a,VanVu2019,VanVu2021,Hasegawa2021,Ishida2025,Nishiyama2025,Prech2025}, yet the field remains scattered: many inequalities, derived by different techniques, address the same core question: how do uncertainty and dissimilarity constrain one another? Entropy production (quantum relative entropy) sometimes appears as the central dissimilarity, sometimes not at all; when it does, the resulting TUR may be suboptimal. This patchwork stems from the absence of a unified toolkit for comparing quantum uncertainty and dissimilarities.

We propose a solution for this situation in two steps. \emph{First}, identify the \emph{fundamental} dissimilarity that natively encodes the uncertainty at hand. In this case, the signal-to-noise ratio. \emph{Second}, represent all other dissimilarities through this fundamental one, so that the fundamental bound is universally inherited.

\emph{Step 1 (Fundamental dissimilarity).} Uncertainty (quantified as variance versus bias) is fundamentally captured by the Pearson $\chi^2$ via the Hammersley-Chapman–Robbins variational identity \cite{Hammersley1950,ChapmanRobbins1951}. For probability distributions $P\ll Q$ and $\Theta \in L^2(Q)$ (${Var}_{Q}(\Theta) < \infty$),
\begin{equation}
\chi^2(P\!\parallel\!Q)
\;=\;
\sup_{\Theta \in L^2(Q)}
\frac{\big|\langle \Theta \rangle_{P}-\langle \Theta \rangle_{Q}\big|^{2}}
{\mathrm{Var}_{Q}(\Theta)},
\label{eq:CR-identity}
\end{equation}
with $\chi^2(P\!\parallel\!Q):=\sum_i (P_i-Q_i)^2/Q_i$. Thus $\chi^2$ is \emph{the} dissimilarity whose value equals the best achievable inverse uncertainty. In that sense, we call it fundamental and it plays a role in the next step.

\emph{Step 2 (Universal lifting).} We lift this backbone to quantum statistics through Petz $f$-divergences \cite{Petz1986,Hiai2011} written in terms of left/right multiplication superoperators. For density operators $\rho,\sigma$,
\begin{equation}
D_f(\rho\!\parallel\!\sigma)
\;:=\;
\Tr\!\left\{\sigma^{1/2}\,
f\!\big(L_{\rho}R_{\sigma^{-1}}\big)\sigma^{1/2}\right\},
\label{eq:Petz-LR}
\end{equation}
where $f:[0,\infty)\to\mathbb{R}$ is continuous in $(0,\infty)$ and $\lim_{x\rightarrow \infty }f(x)/x$ exists in $\mathbb{R}$. We also require $f$ to be operator convex \cite{Hiai2011} ($f(tA+(1-t)B)\leq tf(A)+(1-t)f(B)$, for any positive semi-definite operators $A,B$ and $t \in [0,1]$). The left/right superoperators are defined as $L_\rho(A)=\rho A$, $R_{\sigma^{-1}}(A)=A\sigma^{-1}$ ($\sigma^{-1}$ is the generalized inverse). Our first main result is the \emph{$\chi^2_\lambda$ integral representation}
\begin{equation}
D_f(\rho\!\parallel\!\sigma)
\;=\;\int_{0}^{1} w_f(\lambda)\,\chi^2_{\lambda}(\rho\!\parallel\!\sigma)\,d\lambda,
\label{eq:q-atomic}
\end{equation}
with nonnegative weights $w_f(\lambda) \geq 0$ determined by $f$ (explicit form given later). Here the quantum $\chi^2$–kernel is the following $f$-divergence, 
\begin{equation}
\chi^2_{\lambda}(\rho||\sigma):=D_{f_\lambda}(\rho||\sigma), \qquad f_\lambda(u):=\frac{(u-1)^2}{(1-\lambda)u+\lambda},
\label{chi2_lambda_def}
\end{equation}
for $\lambda \in (0,1)$. Equation~\eqref{eq:q-atomic} states that every Petz $f-$ divergence is a positive mixture of the same quadratic contrasts; different notions of distinguishability (relative entropy, Hellinger/Bures, Petz-Rényi) differ only by the mixing law $w_f(\lambda)$, which selects forward-, backward-, or symmetrically-biased fluctuation sectors.

Moreover, we show \eqref{chi2_lambda_def} is also given in terms of the classical Pearson, $\chi^2$ \eqref{eq:CR-identity} as
$\chi^2_{\lambda}(\rho||\sigma)=(1/\lambda^2)
\chi^2\!\big(P\|(1-\lambda)P+\lambda Q\big)$, where $P,Q$ are the Nussbaum–Szkoła (NS) distributions \cite{Nussbaum2009,Androulakis2023a,Salazar2023d,Androulakis2024} uniquely defined from $\rho$ and $\sigma$ (more details later). Then, combining \eqref{eq:CR-identity} with \eqref{eq:q-atomic} immediately yields a wide family of quantum $f$-divergence TURs. For any Hermitian observable $\hat\theta$ with known $x:=\langle \hat\theta\rangle_{\rho}-\langle \hat\theta\rangle_{\sigma}$, $y:=\mathrm{Var}_{\rho}(\hat\theta)$ and $z:=\mathrm{Var}_{\sigma}(\hat\theta)$, our second main result is the \emph{Universal TUR for quantum (Petz $f$-) divergences}
\begin{equation}
D_f(\rho\!\parallel\!\sigma)
\;\ge\;D_f(\rho_\star||\sigma_\star) =\int_{0}^{1} w_f(\lambda)\,h_{\lambda}(x,y,z)\,d\lambda
\label{eq:f-TUR}
\end{equation}
with the closed-form contrast
\begin{equation}
h_{\lambda}(x,y,z)
\;:=\;
\frac{x^{2}}
{(1-\lambda)\,y+\lambda\,z+\lambda(1-\lambda)\,x^{2}}.
\label{eq:h-lambda}
\end{equation}
Inequality~\eqref{eq:f-TUR} upgrades the integral representation \eqref{eq:q-atomic} into an operational constraint: achieving bias $x$ against the forward/backward variances $(y,z)$ requires spending an $f$-divergence budget. Note that the bound (\ref{eq:f-TUR}) is tight. In this setup (given $x,y,z$), it means that any other bound must be equivalent or weaker than this one. The lower bound is attained by a commuting ($[\rho_\star,\sigma_\star]=0$) binary two level system (details below).

Several applications follow from (\ref{eq:f-TUR}) for choices of $f$. For instance, it recovers the thermodynamic uncertainty relation for the entropy production \cite{Salazar2024d} for the KL divergence; the symmetric quantum relative entropy (Jeffreys) \cite{Salazar2023d} and symmetric Petz-Rényi cases \cite{Salazar2024b}; the denerate $\chi^2$ case \cite{Hasegawa2025a}; and their underlying counterparts in stochastic thermodynamics; as well as new ones, such as the Petz-Rényi (nonsymmetric) and Hellinger/Bures case discussed below.

\textbf{Formalism.—}
We proceed from the classical case to the quantum one via the Nussbaum–Szkoła (NS) bridge \cite{Nussbaum2009,Androulakis2023a,Salazar2023d,Androulakis2024}.

\emph{Classical $f$-divergence and its $\chi^2_\lambda$ integral representation.}
For probability distributions $P,Q$ on a finite (or countable) set with $P\ll Q$, the Csiszár $f$-divergence is
\[
D_f^{\mathrm c}(P\!\parallel\!Q)\;:=\;\sum_i Q_i\,f\!\Big(\frac{P_i}{Q_i}\Big),
\]
for convex $f:(0,\infty)\to\mathbb{R}$. For specific choices of $f(t)$, one gets Kullback-Leibler [$f(t)=t\ln t$], Pearson $\chi^2$ [$f(t)=(t-1)^2$], Total variation [$f(t)=(1/2)|t-1|$] and so on. 
Now we introduce the $\lambda$–mixture $Q_\lambda:=(1-\lambda)P+\lambda Q$ and the classical kernels ($\lambda \in (0,1)$),
\begin{equation}
\chi_\lambda(P||Q):=\;\frac{1}{\lambda^2}\chi^2\!\big(P\!\parallel\!Q_\lambda\big)
=\frac{1}{\lambda^2}\sum_i\frac{(P_i-Q_{\lambda,i})^2}{Q_{\lambda,i}},
\label{classic_chi2}
\end{equation}
and define
\begin{equation}
g(u)\;:=\;\frac{f(u)}{(u-1)^2},\qquad u>0.
\label{g_u}
\end{equation}
A classical result (Löwner/Pick–Nevanlinna representation) states that, for a broad class of generators $f$ (called operator convex \cite{Hiai2011}), $g$ is a \emph{Stieltjes function}, i.e.
\begin{equation}
g(u)\;=\;\int_{0}^{\infty}\frac{\phi_f(t)}{\,1+t\,u\,}\,dt,\qquad \phi_f(t)\ge 0,
\label{eq:stieltjes}
\end{equation}
with $\phi_f$ determined uniquely by $f$. Plugging \eqref{eq:stieltjes} back gives
\begin{equation}
D_f^c(P\|Q)
=\int_{0}^{\infty}\!\phi_f(t)\,\sum_x q_x\frac{(u_x-1)^2}{1+t u_x}\,dt.
\label{eq:csiszar-chi}
\end{equation}
with change variables $t=\tfrac{1-\lambda}{\lambda}$ and
$d\lambda=-\tfrac{1}{(1+t)^2}dt$). A short algebraic match yields
\begin{equation}
D_f^c(P\|Q) \;=\; \int_0^1 w_f(\lambda)\,\chi^2_\lambda(P\|Q)\,d\lambda ,
\label{eq:csiszar-chi}
\end{equation}
with a nonnegative weight (see properties further in the SM),
\begin{equation}
w_f(\lambda)\;=\;\frac{1}{\lambda}\;
\phi_f\!\Big(\frac{1-\lambda}{\lambda}\Big),\qquad \lambda\in(0,1).
\label{eq:w-from-phi}
\end{equation}
In words: \emph{$\chi^2_\lambda$ are the nonnegative spectral weights any $f$-divergence is a positive mixture of them.} A practical route to get $w_f(\lambda)$ is
using Stieltjes inversion (directly from $f$).
Compute $g(u)$ from (\ref{g_u}). The inversion formula gives (See SM), for $t>0$,
\begin{equation}
\phi_f(t)\;=\;\frac{1}{\pi t}\,\lim_{\varepsilon\downarrow 0}
\operatorname{Im}\, g\!\Big(-\frac{1}{t}-i\varepsilon\Big),
\label{eq:stieltjes-inversion}
\end{equation}
and then \eqref{eq:w-from-phi} yields $w_f$ exactly. In summary, from a given $f$, one defines $g$ using (\ref{g_u}), then finds $\phi_f$ using (\ref{eq:stieltjes-inversion}) and $w_\lambda$ using (\ref{eq:w-from-phi}).

\emph{Quantum lifting via NS.}
Consider density matrices $\rho$ and $\sigma$ with spectral decompositions $\rho=\sum_i p_i|p_i\rangle\!\langle p_i|$ and $\sigma=\sum_j q_j|q_j\rangle\!\langle q_j|$, and define the NS pair
\[
P_\rho(i,j):=p_i|\langle p_i|q_j\rangle|^2,\qquad
Q_\sigma(i,j):=q_j|\langle p_i|q_j\rangle|^2.
\]
For the Petz $f$-divergence in LR form ~\eqref{eq:Petz-LR}, the NS correspondence matches the quantum divergence from a theory in quantum information (see SM), 
\begin{equation}
D_f(\rho\!\parallel\!\sigma)=D_f^{\mathrm c}(P_\rho\!\parallel\!Q_\sigma).
\label{eq:NS}
\end{equation}
Combining \eqref{eq:csiszar-chi} with \eqref{eq:NS} yields the \emph{quantum $\chi^2_\lambda$ integral representation} ~\eqref{eq:q-atomic}, with the quantum kernel \eqref{chi2_lambda_def} given explicitly
\begin{equation}
\chi^2_\lambda(\rho||\sigma)=\sum_{ij}\frac{(p_i-q_j)^2}{(1-\lambda)p_i+\lambda q_j}|\langle p_i|q_j\rangle|^2.
\label{chi2:explicit}
\end{equation}

\emph{From kernels to TURs.}
Let $\hat\theta$ be Hermitian with finite variances under both states and set 
$x:=\langle\hat\theta\rangle_\rho-\langle\hat\theta\rangle_\sigma$, 
$y:=\mathrm{Var}_\rho(\hat\theta)$, 
$z:=\mathrm{Var}_\sigma(\hat\theta)$. Extend the NS distribution to operators \cite{Salazar2023d} by defining the (complex) random variable $\Theta(i,j)=\langle p_i|\hat \theta |q_j \rangle/\langle p_i|q_j\rangle$, for $\langle p_i|q_j\rangle\neq0$, and $\Theta(i,j)=0$. We have $X:=Var_P(\Theta)\leq x$, $Y:=Var_Q(\Theta)\leq y$ and $Z=|E_P(\Theta)-E_Q(\Theta)|=z$.
Applying the Chapman–Robbins variational identity ~\eqref{eq:CR-identity} in \eqref{classic_chi2} gives
\begin{equation}
\chi^2_\lambda(\rho\!\parallel\!\sigma)=\lambda^{-2}\chi^2(P||Q_\lambda)\geq h_\lambda(X,Y,Z)\ge h_\lambda(x,y,z),
\label{eq:CR}
\end{equation}
with $h_\lambda$ as in \eqref{eq:h-lambda}, which is decreasing in the arguments $y,z$. Substituting \eqref{eq:CR} into the quantum integral representation \eqref{eq:q-atomic} yields the $f$-divergence TUR ~\eqref{eq:f-TUR}. The weight $w_f$ selects which fluctuation sectors dominate a given $D_f$, transporting the Chapman–Robbins precision constraint from $\chi^2_\lambda$ to any Petz $f-$divergence. In the SM, we prove a stronger version, where it shows that the lower bound (\ref{eq:f-TUR}) is tight, $D_f(\rho||\sigma)\geq D_f(\rho_\star||
\sigma_\star )$, with $\rho_\star=r |u_1 \rangle \langle u_1| + (1-r)|u_2\rangle\langle u_2|$ and $\sigma_\star=s |u_1 \rangle \langle u_1| + (1-s)|u_2\rangle\langle u_2|$, where $|u_1\rangle, |u_2\rangle$ are eigenvectors of $\hat \theta$ (with eigenvalues $u_1 \neq u_2$) and $r,s$ are functions of $(x,y,z)$ given as $r=\tfrac12+\frac{b+\Delta^2}{4\Delta v}$, $s=\tfrac12+\frac{b-\Delta^2}{4\Delta v}$, $\Delta:=x$, $b:=z-y$, $v:=\frac{1}{2|\Delta|}\sqrt{b^2+2\Delta^2(y+z)+\Delta^4}$, for which we get $\chi^2_\lambda(\rho_\star||\sigma_\star)=h_\lambda(x,y,z)$ and
\begin{equation}
D_f(\rho_\star||\sigma_\star)=\int_0^1 w_f(\lambda)h_\lambda(x,y,z)d\lambda.
\end{equation}

\textbf{Properties of the \texorpdfstring{$\chi^2_\lambda$}{chi2-lambda} integral representation—} The $\chi^2_\lambda$ integral representation in Eqs.~\eqref{eq:csiszar-chi} and \eqref{eq:q-atomic}, with mixing law $w_f(\lambda)$ given by \eqref{eq:w-from-phi}, endows each $f$-divergence with a transparent “weight” structure over fluctuation sectors indexed by $\lambda\in(0,1)$. The items below summarize the key structural facts we use throughout; proofs and technicalities appear in the Supplementary Material (SM).

\paragraph*{(P1) Nonnegativity and uniqueness}
For any admissible $f$, the weight is nonnegative: 
\begin{equation}
w_f(\lambda)\ge 0,
\end{equation}
on $(0,1)$, and it is unique almost everywhere (a.e.). This property was particularly useful to derive our universal quantum TUR (\ref{eq:f-TUR}).

\paragraph*{(P2) Linearity and affine invariance.}
The map $f\mapsto w_f$ is linear on the cone of convex generators, and it is invariant under adding affine pieces:
\begin{equation}
w_{\alpha f_1+f_2}=\alpha\,w_{f_1}+w_{f_2},\qquad
w_{\,f+a+b(u-1)}=w_f.
\label{Linearityw}
\end{equation}
Physically, affine shifts do not change $D_f^{\mathrm c}$ (nor $D_f$) and hence do not alter the mixing law.

\paragraph*{(P3) Order preservation.}
If $f_1\le f_2$ pointwise (modulo affine terms), then $w_{f_2}-w_{f_1}\ge 0$ a.e. Thus the partial order on divergences is mirrored by a partial order on the weights.

\paragraph*{(P4) Endpoints of $f$.}
The weight encodes key boundary/curvature data of $f$. After harmless centering at $u=1$, for differentiable $f$, set $\tilde f(u)=f(u)-f(1)-f'(1)(u-1)$, one has
\begin{subequations}\label{eq:moment-relations}
\begin{align}
\tilde f(0^+)&=\int_0^1 \lambda^{-1}\,w_f(\lambda)\,d\lambda, \label{eq:moment-relations-a}\\
\lim_{u\to\infty}\frac{f(u)}{u}&=\int_0^1 (1-\lambda)^{-1}\,w_f(\lambda)\,d\lambda.
\label{eq:moment-relations-b}
\end{align}
\end{subequations}

\paragraph*{(P5) Duality under inversion.}
For the dual generator $f^\ast(u):=u\,f(1/u)$ one has \text{a.e. on }(0,1)
\begin{equation}
w_{f^\ast}(\mu)=\,w_f(1-\mu).
\label{dualw}
\end{equation}
Thus exchanging forward/backward sectors ($\lambda\leftrightarrow 1-\lambda$) corresponds to the standard $f\!\leftrightarrow\! f^\ast$ duality (Jeffreys/symmetric cases inherit a self-consistency under this map).

\paragraph*{(P6) Small-contrast control and all orders.}
Let $\tilde f(u):=f(u)-f(1)-f'(1)(u-1)$ so that $\tilde f(1)=\tilde f'(1)=0$. Then, for $|\varepsilon|<1$,
\begin{equation}
\label{SM:smallcontrast}
\tilde f(1+\varepsilon)
=\sum_{n=2}^{\infty}(-1)^{n}\!\left(\int_{0}^{1}\lambda^{n}\,w_f(\lambda)\,d\lambda\right)\varepsilon^{n},
\end{equation}
and, equivalently, for each $n\ge 2$,
\begin{equation}
\label{SM:moment-identity}
\frac{\tilde f^{(n)}(1)}{n!}
= (-1)^{n-2}\!\int_{0}^{1}(1-\lambda)^{n-2}\,w_f(\lambda)\,d\lambda.
\end{equation}
Particularly, we have $\tilde f''(1)/2 = \int_0^1 w_f(\lambda)d\lambda > 0$, which ensures that $w_f(\lambda)$ is normalizable.

\textbf{Application: Quantum Thermodynamic Uncertainty Relations—}
We illustrate how the integral representation \eqref{eq:q-atomic} and the universal TUR \eqref{eq:f-TUR} combine into a concrete strategy that turns a generator $f$ (operator convex) into an explicit ``thermodynamic uncertainty relation''. Consider the usual setup in quantum thermodynamics \cite{LandiReview2021}, where we have system + environment evolving with a unitary ($U$), the entropy production is defined with respect to the following states: $\rho:=\rho_{SE}'$ and $\sigma:=\rho_S'\otimes\rho_E$, where $\rho_{SE}':=U(\rho_S \otimes \rho_E) U^\dagger$ and $\rho_S':=\Tr_E(\rho_{SE}')$. In this case, we have the entropy production $\Sigma:=D(\rho||\sigma)$. Additionally, let $\hat\theta$ be an Hermitian operator and $x:=\langle \hat\theta\rangle_\rho - \langle \hat\theta\rangle_\sigma$, $y:={\rm Var}_\rho(\hat\theta)$, $z:={\rm Var}_\sigma(\hat\theta)$. In what follows, for each case we (i) identify the generator $f$, (ii) recall the familiar closed form of $D_f(\rho||\sigma)$, (iii) retrieve the mixing weight $w_f(\lambda)$ from the Sup. Mat. (SM), and (iv) write the resulting TUR in a form that highlights structure or simplifications. In all cases, the resulting integral is given by $D_f(\rho_\star|\sigma_\star)$ from \eqref{eq:f-TUR}, but we choose to display the lower bound as an integral of $h_\lambda$ in order to understand how TURs differ only by their spectral weights $w_f(\lambda)$.

\emph{Quantum relative entropy (Entropy Production).}
The forward KL is generated by $f(u)=u\log u-(u-1)$, yielding the standard expression
$D(\rho||\sigma)=\Tr[\rho(\log\rho-\log\sigma)]$.
The SM provides the weight (via \eqref{eq:w-from-phi}) as
\(
w_{\mathrm{KL}}(\lambda)=\lambda
\).
Substituting in \eqref{eq:f-TUR} we obtain
\begin{equation}
\Sigma=D(\rho\!\parallel\!\sigma)\ \ge\
\int_0^1 \lambda\,
h_\lambda(x,y,z)\,d\lambda.
\end{equation}
This is the TUR for the quantum entropy production \cite{Salazar2024d}, as expected.

\emph{Reverse quantum relative entropy.}
For the reverse KL, $f(u)=-\log u+(u-1)$ and
$D_{\rm rKL}(\rho\!\parallel\!\sigma)=D(\sigma||\rho)$, we use property (\ref{dualw}), $w_{\mathrm{rKL}}(\lambda)=w_{KL}(1-\lambda)=(1-\lambda)$, 
so that
\begin{equation}
D(\sigma||\rho)\ \ge\
\int_0^1 (1-\lambda)
h_\lambda(x,y,z)\,d\lambda,
\end{equation}
which is immediate also from the symmetries of $h_\lambda(x,y,z)=h_{1-\lambda}(-x,z,y)$.

\emph{Jeffreys divergence.} Symmetric quantum relative entropy (also known as Jeffreys divergence) is given by $D(\rho,\sigma)=(1/2)(D(\rho||\sigma)+D(\sigma||\rho))$. Using property (\ref{Linearityw}), we have $w_f(\lambda)=(1/2)[w_{KL}(\lambda)+w_{rKL}(\lambda)]=1$. Thus, the Quantum TUR,

\begin{equation}
D(\sigma,\rho)\ \ge\
\int_0^1 
h_\lambda(x,y,z)\,d\lambda,
\end{equation}
which is also equivalent to a previous result \cite{Salazar2023d} and its classical equivalent, sometimes called exchange TUR \cite{Timpanaro2019B}.

\emph{Pearson $\chi^2$.}
As a sanity check, for $f(u)=(u-1)^2$ one has the Pearson divergence
$D_{\mathrm P}(\rho\!\parallel\!\sigma)=\chi^2_{\,1}=\chi^2(\rho\!\parallel\!\sigma)$ \eqref{classic_chi2}.
The SM shows that the mixture reduces to a point mass,
\(
w_{\mathrm P}(\lambda)=\delta(1-\lambda)
\),
so \eqref{eq:f-TUR} collapses to a (quantum) Chappman-Robbins relation,
\begin{equation}
\chi^2(\rho\!\parallel\!\sigma)\ \ge\ h_{1}(x,y,z)=\frac{x^2}{z},
\end{equation}
which is also written as
\begin{equation}
\frac{\langle\langle\hat \theta\rangle\rangle_\sigma}{(\langle \hat\theta \rangle _\rho -\langle \hat \theta \rangle_\sigma )^2 }\geq  \frac{1}{\Tr(\rho^2 \sigma^{-1})-1},,
\label{eq:apps-pearson-ur}
\end{equation}
after using $\chi^2_1(\rho||\sigma )=\Tr(\rho^2 \sigma^{-1})-1$. Equation (\ref{eq:apps-pearson-ur}) is one step away from a previous result \cite{Hasegawa2025a}, where they further optimize over all possible $\sigma$.

\emph{Squared Hellinger.}
For the Hellinger generator $f(u)=(1/2)(\sqrt{u}-1)^2$, the divergence equals
$H^2(\rho,\sigma)=\big(1-\Tr[\sqrt{\rho}\sqrt{\sigma}]\big)$ in the Petz framework.
The SM yields the nontrivial algebraic weight
\(
w_{H^2}(\lambda)=\frac{1}{\pi}\sqrt{\lambda(1-\lambda)}
\) (See SM).
Consequently from (\ref{eq:f-TUR}), we have
\begin{equation}
H^2(\rho,\sigma)\ \ge\
1-\Bigg(1+\frac{(\langle \hat\theta)_\rho-\langle \hat\theta\rangle_\sigma)^2}{(\sqrt{\langle \langle\hat\theta\rangle\rangle_\rho}+\sqrt{\langle \langle\hat\theta\rangle\rangle_\sigma})^2}\Bigg)^{-1/2},
\end{equation}
which reproduces a previous classical result also used in the context of quantum thermodynamics \cite{Nishiyama2025b}.

\emph{Petz--Rényi} 
This is a novel contribution of the Universal TUR and this case is related to  cumulant generating functions \cite{LandiReview2021}, therefore the integral representation is valuable to understand the statistics of operators. With
\(
f_\alpha(u)=\frac{u^\alpha-1-\alpha(u-1)}{\alpha(1-\alpha)}
\),
for $\alpha\!\in\!(0,1)$, which guarantees $f$ is operator convex, and the divergence reads
\(
D_\alpha(\rho\!\parallel\!\sigma)
:=\frac{1}{\alpha(1-\alpha)}\big(\Tr[\rho^\alpha\sigma^{1-\alpha}]-1\big)
\).
We obtain
\[
w_\alpha(\lambda)
=\frac{\sin(\pi\alpha)}{\pi\,\alpha(1-\alpha)}\,
\lambda^{\alpha}(1-\lambda)^{1-\alpha}
\geq 0\]
so
\begin{equation}
D_\alpha(\rho\!\parallel\!\sigma)\ \ge\
\frac{\sin(\pi\alpha)}{\pi\,\alpha(1-\alpha)}\int_0^1
\lambda^{\alpha}(1-\lambda)^{1-\alpha}\,
h_\lambda(x,y,z)\,d\lambda.
\label{alphaTUR}
\end{equation}
As $\alpha\uparrow 1$ the weight tends to $w_0(\lambda)=\lambda$ (KL), as $\alpha \downarrow0$, it tends to $w_0(\lambda)=1-\lambda$ (rKL) and $w_{1/2}(\lambda)=4w_{H^2}(\lambda)$ (Hellinger's). And the symmetric $\alpha$ divergence is related to a symmetric TUR \cite{Salazar2024b} for $\alpha \in (0,1)$.

In quantum thermodynamics, the $\alpha-$Divergence appears naturally as a cummulant generating function (CGF) of the nonequilibrim lag problem \cite{LandiReview2021,Campisi2011}. In this context, a system S undergoes a work protocol starting from thermal equilibrium state, $\rho_i^{th}=e^{-\beta H_i}/Z_i$, and evolves in time driven by unitary generated by the time dependent Hamiltonian $H(t)$ ($H(0)=H_i$, $H(\tau)=H_f$), $\rho'=V\rho^{th}_iV^\dagger$. After a time $\tau$, the system thermalizes from the nonequilibrium state $\rho'$ to $\rho^{th}_f=e^{-\beta H_f}/Z_f$. The entropy production is given by $\Sigma=D(\rho'||\rho_f^{th})$ and it is also called the nonequilibrum lag. The nonequilibrium CDF \cite{Talkner2007,Esposito2009,Guarnieri2019} (which is related to statistics of a random variable $S$ such that $\langle S \rangle=\Sigma$) is given by
\begin{equation}
K(\alpha)=\ln \Tr[(\rho_f^{th})^\alpha (\rho')^{1-\alpha}].
\end{equation}
From our definition of $D_\alpha$, we see that $K(\alpha)=\ln[1+\alpha(1-\alpha)D_\alpha(\rho_f^{th}||\rho')]$ and from (\ref{alphaTUR}) we obtain a lower bound for the CDF
\begin{equation}
K(\alpha)\geq \ln\bigg[1+\frac{\sin(\pi\alpha)}{\pi}\int_0^1\lambda^\alpha (1-\lambda)^{1-\alpha} h_\lambda(x,y,z)d\lambda\bigg],
\end{equation}
for $\alpha \in (0,1)$, in terms of any system observable $\hat \theta$ with $x=\langle \hat\theta\rangle_{\rho^{th}_f}-\langle \hat\theta\rangle_{\rho'}$, $y=\mathrm{Var}_{\rho^{th}_f}(\hat\theta)$ and $z=\mathrm{Var}_{\rho'}(\hat\theta)$.

\emph{Triangular discrimination.}
For
\(
f(u)=\frac{(u-1)^2}{u+1}
\),
the triangular discrimination obeys the quantum identity
\(
\Delta(\rho,\sigma)=(1/2)\,\chi^2_{1/2}(\rho||\sigma)
\)
, with $w_\Delta(\lambda)=(1/2)\delta(\lambda-1/2)$, which results in from (\ref{eq:f-TUR})
\begin{equation}
\Delta(\rho,\sigma)\ \ge\ \frac{1}{2}\,h_{1/2}(x,y,z)
=\frac{x^{2}}{\,y+z+ x^{2}/2\,}.
\end{equation}
The mid-mixture slice $\lambda=\tfrac12$ gives a balanced denominator, tying the TUR directly to a symmetric interpolation between $y$ and $z$, yielding the following TUR
\begin{equation}
\frac{\langle \langle \hat \theta \rangle \rangle_\rho + \langle \langle \hat\theta \rangle\rangle_\sigma}{(1/2)(\langle \hat\theta \rangle_\rho - \langle \hat\theta\rangle_\sigma)^2 } \geq \frac{1-\Delta(\rho,\sigma)/2}{\Delta(\rho,\sigma)/2},
\end{equation}
also previously obtained in the context of quantum information \cite{Salazar2024d}.

\textbf{Acknowledgment—} During the preparation of this work the author used GPT-5 (OpenAI) in order to proofread the manuscript and enhance mathematical proofs. After using this tool, the author reviewed and edited the content as needed and takes full responsibility for the content of the published article.

\textbf{Conclusions—}
We introduced a two–step, broadly applicable strategy to derive sharp, information–theoretic uncertainty relations for quantum thermodynamics. The method is modular: a \emph{variational backbone} that captures the operational uncertainty of interest, coupled with an \emph{integral lifting} that transports that backbone to full quantum Petz $f$–divergences.

The starting point is the variational representation of the classical $\chi^2$–divergence, the Chapman–Robbins identity in Eq.~\eqref{eq:CR-identity}. Through the Nussbaum–Szkoła distributions, this leads to the quantum TUR (\ref{eq:f-TUR}). This step is problem–driven: in this work the relevant quantum uncertainty is a mean–square to variance ratio, so $\chi^2$ is the \emph{fundamental dissimilarity} that matches the structure of the task. For different uncertainty functionals (e.g., constraints built from higher moments, skewness–like functionals, or local susceptibilities), one should swap in the variational identity that naturally encodes the corresponding control–to–noise tradeoff, thereby replacing $h_\lambda$ by the contrast induced by the new backbone.
We showed our strategy resulted in several known and novel results in quantum thermodynamics. Namely, the quantum entropy production TUR, the symmetric quantum relative entropy TUR, a lower bound for the CDF in the nonequilibrium lag problem and bonds for other dissimilarities (triangular discrimination, Hellinger's). We expect the result will be useful and extended to unravel new properties of systems in quantum thermodynamics.

\bibliography{library}

\clearpage
\onecolumngrid
\begin{center}
\textbf{\large Supplemental Material: Universal Thermodynamic Uncertainty Relation for Quantum $f-$Divergences}
\end{center}
\setcounter{equation}{0}
\renewcommand{\theequation}{S\arabic{equation}}


\section{Nussbaum--Szkoła distributions and equality of divergences}
\textbf{Nussbaum--Szkoła bridge (classical $\leftrightarrow$ quantum).—}
We show that Petz $f$-divergences in the LR–superoperator form (Intro, Eq.~\eqref{eq:Petz-LR}) and the $\chi^2_\lambda$ kernels reduce to their classical counterparts under the Nussbaum–Szkoła (NS) “classicalization.”

\medskip
\emph{Setup.} Let $\rho=\sum_i p_i |p_i\rangle\!\langle p_i|$ and $\sigma=\sum_j q_j |q_j\rangle\!\langle q_j|$ be spectral decompositions, and set
\[
a_{ij}:=|\langle p_i|q_j\rangle|^2,\qquad \sum_j a_{ij}=\sum_i a_{ij}=1.
\]
Define the NS distributions
\[
P_\rho(i,j):=p_i\,a_{ij},\qquad Q_\sigma(i,j):=q_j\,a_{ij}.
\]
Write the left/right multiplication superoperators as $L_\rho(X)=\rho X$ and $R_{\sigma^{-1}}(X)=X\sigma^{-1}$.
The Petz divergence used in the \emph{Introduction} is
\[
D_f(\rho\!\parallel\!\sigma)=\operatorname{Tr}\!\Big\{\sigma^{1/2}\,
f\!\big(L_{\rho}R_{\sigma^{-1}}\big)\,
\sigma^{1/2}\Big\},
\]
for admissible $f$ (operator-convex on $(0,\infty)$).

\medskip
\emph{Diagonalizing $L_\rho R_{\sigma^{-1}}$.}
Consider rank-one operators $E_{ij}:=|p_i\rangle\!\langle q_j|$. Then
\[
(L_\rho R_{\sigma^{-1}})(E_{ij})=\rho E_{ij}\sigma^{-1}=\frac{p_i}{q_j}\,E_{ij},
\]
so $\{E_{ij}\}$ diagonalizes $L_\rho R_{\sigma^{-1}}$ with eigenvalues $p_i/q_j$ and
$f(L_\rho R_{\sigma^{-1}})E_{ij}=f(p_i/q_j)E_{ij}$.

\medskip
\emph{Equality of $f$-divergences.}
Expand $\sigma^{1/2}=\sum_j \sqrt{q_j}\,|q_j\rangle\!\langle q_j|
=\sum_{i,j}\sqrt{q_j}\,\langle p_i|q_j\rangle\,E_{ij}$ and compute
\[
\begin{aligned}
D_f(\rho\!\parallel\!\sigma)
&=\big\langle \sigma^{1/2},\, f(L_\rho R_{\sigma^{-1}})\,\sigma^{1/2}\big\rangle_{HS}\\
&=\sum_{i,j,k,\ell}\sqrt{q_j q_\ell}\,\langle p_i|q_j\rangle\,\overline{\langle p_k|q_\ell\rangle}\,
\big\langle E_{ij},\, f(L_\rho R_{\sigma^{-1}})E_{k\ell}\big\rangle_{HS}\\
&=\sum_{i,j} q_j\,|\langle p_i|q_j\rangle|^2\, f\!\Big(\frac{p_i}{q_j}\Big)
=\sum_{i,j} Q_\sigma(i,j)\, f\!\Big(\frac{P_\rho(i,j)}{Q_\sigma(i,j)}\Big).
\end{aligned}
\]
Hence
\[
\boxed{\,D_f(\rho\!\parallel\!\sigma)=D_f^{\mathrm c}(P_\rho\!\parallel\!Q_\sigma)\,}.
\]

\medskip
\emph{Equality for the $\lambda$–$\chi^2$ kernel.}
In the quantum notation used in the paper (Formalism; Intro), define the resolvent-like superoperator
\[
\mathcal{K}_\lambda:=(1-\lambda)\,L_\rho+\lambda\,R_\sigma,\qquad \lambda\in(0,1),
\]
and the quantum kernel
\[
\chi^2_\lambda(\rho\!\parallel\!\sigma):=\,\operatorname{Tr}\!\Big[(\rho-\sigma)\,\mathcal{K}_\lambda^{-1}(\rho-\sigma)\Big].
\]
On the basis $E_{ij}$ we have $\mathcal{K}_\lambda(E_{ij})=\big((1-\lambda)p_i+\lambda q_j\big)E_{ij}$, so
\[
\chi^2_\lambda(\rho\!\parallel\!\sigma)
=\sum_{i,j}\frac{(p_i-q_j)^2}{(1-\lambda)p_i+\lambda q_j}\,|\langle p_i|q_j\rangle|^2.
\]
Since $P_\rho(i,j)=p_i a_{ij}$ and $Q_\sigma(i,j)=q_j a_{ij}$, this equals
\[
\frac{1}{\lambda^2}\sum_{i,j}\frac{\lambda^2\,(P_\rho(i,j)-Q_\sigma(i,j))^2}{(1-\lambda)P_\rho(i,j)+\lambda Q_\sigma(i,j)}
=\frac{1}{\lambda^2}\chi^2\!\big(P_\rho\,\big\|\,(1-\lambda)P_\rho+\lambda Q_\sigma\big)=\chi^2_\lambda\!\big(P_\rho\,\big\|\,Q_\sigma\big).
\]
Therefore
\[
\boxed{\,\chi^2_\lambda(\rho\!\parallel\!\sigma)=\chi^2_\lambda(P_\rho\!\parallel\!Q_\sigma)\,}.
\]

\medskip
\emph{Conclusion.}
Combining the identities above with the classical $\chi^2_\lambda$ integral representation
$D_f^{\mathrm c}(P\|Q)=\int_0^1 w_f(\lambda)\,\chi^2_\lambda(P\|Q)\,d\lambda$
(Formalism, Eq.~\eqref{eq:csiszar-chi}) yields the quantum mixture
\[
D_f(\rho\!\parallel\!\sigma)=\int_0^1 w_f(\lambda)\,\chi^2_\lambda(\rho\!\parallel\!\sigma)\,d\lambda,
\]
which is Eq.~\eqref{eq:q-atomic} in the \emph{Introduction}/\emph{Formalism}.

\section{$\chi^2_\lambda$–mixture and a universal TUR}
\label{SM(section-merged)}

\noindent\textbf{Standing assumptions and notation.}
We work with Petz $f$–divergences that admit the nonnegative $\chi^2_\lambda$–mixture
\begin{equation}
\label{SM(chi-mix)}
\Df(\,\cdot\,\|\,\cdot\,)=\int_0^1 w_f(\lambda)\,\chi^2_\lambda(\,\cdot\,\|\,\cdot\,)\,d\lambda,
\qquad w_f(\lambda)\ge 0.
\end{equation}
In the \emph{classical} case on a countable alphabet $\Omega=\{u_i\}_{i\in I}$ with pmfs $p_i:=P(u_i)$ and $q_i:=Q(u_i)$, we adopt the \emph{actual} kernel (your convention)
\begin{equation}
\label{SM(chi-classical)}
\chi^2_\lambda(P\|Q)
=\sum_{i\in I}\frac{(p_i-q_i)^2}{\lambda\,q_i+(1-\lambda)\,p_i}\,,
\qquad \lambda\in(0,1).
\end{equation}
In the \emph{quantum} case, with $L_\rho(X)=\rho X$ and $R_\sigma(X)=X\sigma$, define
\begin{equation}
\label{SM(chi-quantum)}
K_\lambda:=(1-\lambda)\,L_\rho+\lambda\,R_\sigma,
\qquad
\chi^2_\lambda(\rho\|\sigma)
=\Tr\!\big[(\rho-\sigma)\,K_\lambda^{-1}(\rho-\sigma)\big].
\end{equation}
(With this choice, the commuting reduction of \eqref{SM(chi-quantum)} exactly reproduces \eqref{SM(chi-classical)}.)

Given a single observable $X$ (classical) or $\hat\theta$ (quantum), define the moment triple
\begin{equation}
\label{SM(x-y-z)}
x:=\mathbb{E}_P[X]-\mathbb{E}_Q[X]\ \ (\text{or }x=\langle\hat\theta\rangle_\rho-\langle\hat\theta\rangle_\sigma),\quad
y:=\mathrm{Var}_P[X]\ \ (\mathrm{Var}_\rho\hat\theta),\quad
z:=\mathrm{Var}_Q[X]\ \ (\mathrm{Var}_\sigma\hat\theta).
\end{equation}
With the kernel in \eqref{SM(chi-classical)}–\eqref{SM(chi-quantum)}, the Chapman–Robbins function becomes
\begin{equation}
\label{SM(h-lambda-def)}
h_\lambda(x,y,z):=\frac{x^2}{(1-\lambda)\,y+\lambda\,z+\lambda(1-\lambda)\,x^2},
\qquad
\lambda\in(0,1),\quad h_0:=0,\ \ h_1:=\frac{x^2}{z}.
\end{equation}

\medskip
\noindent\textbf{Lemma.\ref{SM(CR)}} (Chapman–Robbins).
\emph{Classical:} For any statistic $\theta$,
\begin{equation}
\label{SM(CR-classical)}
\chi^2_\lambda(P\|Q)\ \ge\ h_\lambda\!\big(x,y,z\big).
\end{equation}
\emph{Quantum:} For any observable $\hat\theta$,
\begin{equation}
\label{SM(CR-quantum)}
\chi^2_\lambda(\rho\|\sigma)\ \ge\ h_\lambda\!\Big(\langle\hat\theta\rangle_\rho-\langle\hat\theta\rangle_\sigma,\ \mathrm{Var}_\rho\hat\theta,\ \mathrm{Var}_\sigma\hat\theta\Big).
\end{equation}

\begin{proof}[Proof (classical)]
Let $\bar\mu_\lambda:=(1-\lambda)\,P+\lambda\,Q$ and 
\[
\bar r_\lambda:=\frac{(p-q)}{\lambda\,q+(1-\lambda)\,p}\,.
\]
Then
\[
\chi^2_\lambda(P\|Q)=\sum_i \bar\mu_{\lambda,i}\,\bar r_{\lambda,i}^2=\|\bar r_\lambda\|_{L^2(\bar\mu_\lambda)}^2,
\qquad
\sum_i \bar\mu_{\lambda,i}\,\bar r_{\lambda,i}=\sum_i (p_i-q_i)=0.
\]
For any $c\in\mathbb{R}$,
\[
x=\sum_i (p_i-q_i)\theta_i=\sum_i \bar\mu_{\lambda,i}\,\bar r_{\lambda,i},(\theta_i-c).
\]
Cauchy–Schwarz gives
\[
x^2\le \Big(\sum_i\bar\mu_{\lambda,i}\bar r_{\lambda,i}^2\Big)\Big(\sum_i\bar\mu_{\lambda,i}(\theta_i-c)^2\Big).
\]
Minimizing over $c$ at $c^\star=\mathbb{E}_{\bar\mu_\lambda}[\theta]=(1-\lambda)\,\mathbb{E}_P[\theta]+\lambda\,\mathbb{E}_Q[\theta]$ and using the mixture-variance identity
\[
\sum_i\bar\mu_{\lambda,i}(\theta_i-c^\star)^2=(1-\lambda)\,y+\lambda\,z+\lambda(1-\lambda)\,x^2
\]
yields \eqref{SM(CR-classical)} with $h_\lambda$ as in \eqref{SM(h-lambda-def)}.
\end{proof}

\begin{proof}[Proof sketch (quantum)]
Let $(P,Q)$ be the Nussbaum–Szkoła pair for $(\rho,\sigma)$. With 
$\theta(i,j):=\langle p_i|\hat\theta|q_j\rangle/\langle p_i|q_j\rangle$, if $\langle p_i | q_j\rangle \neq 0$ (and $=0$, otherwise) one has
$\mathbb{E}_P[\theta]=\langle\hat\theta\rangle_\rho$, $\mathbb{E}_Q[\theta]=\langle\hat\theta\rangle_\sigma$, and
$\mathrm{Var}_P(\theta)\le \mathrm{Var}_\rho\hat\theta$, $\mathrm{Var}_Q(\theta)\le \mathrm{Var}_\sigma\hat\theta$ (pinching).
Moreover, the Petz–$\chi^2_\lambda$ in \eqref{SM(chi-quantum)} reduces to the classical expression \eqref{SM(chi-classical)} on the NS pair. Since $h_\lambda$ is decreasing in each of $y,z$, the classical bound transfers to \eqref{SM(CR-quantum)}.
\end{proof}
\label{SM(CR)}

\medskip
\noindent\textbf{Binary parametrization and saturation.}
Given $(x,y,z)$ with $y,z\ge 0$ and $x\ne 0$, set
\begin{equation}
\label{SM(v-def)}
\Delta:=x,\qquad b:=z-y,\qquad
v:=\frac{1}{2|\Delta|}\sqrt{b^2+2\Delta^2(y+z)+\Delta^4},\qquad
r=\tfrac12+\frac{b+\Delta^2}{4\Delta v},\quad s=\tfrac12+\frac{b-\Delta^2}{4\Delta v}.
\end{equation}
Then for all $\lambda\in(0,1)$,
\begin{equation}
\label{SM(chi-equals-h)}
\chi^2_\lambda\!\big(\mathrm{Bern}(r)\,\big\|\,\mathrm{Bern}(s)\big)=h_\lambda(x,y,z).
\end{equation}
In the quantum setting, if $\hat\theta$ has eigenvectors $\{|u_1\rangle,|u_2\rangle\}$ with eigenvalues $a<b$ realizing $(x,y,z)$, the commuting qubit states
\begin{equation}
\label{SM(sat:states)}
\rho_\star=r\,|u_1\rangle\!\langle u_1|+(1-r)\,|u_2\rangle\!\langle u_2|,\qquad
\sigma_\star=s\,|u_1\rangle\!\langle u_1|+(1-s)\,|u_2\rangle\!\langle u_2|
\end{equation}
satisfy
\begin{equation}
\label{SM(sat:mm)}
r u_1+(1-r)u_2=\langle\hat\theta\rangle_\rho,\quad s u_1+(1-s)u_2=\langle\hat\theta\rangle_\sigma,\quad
(u_2-u_1)^2=\frac{y}{r(1-r)}=\frac{z}{s(1-s)}.
\end{equation}

\bigskip
\noindent\textbf{Proposition \ref{SM(prop-binary)}} (Binary optimality via $\chi^2_\lambda$–mixture).
\emph{(a) Classical.}
For any feasible $(x,y,z)$,
\begin{equation}
\label{SM(inf-classical)}
\inf_{(P,Q):\,(\ref{SM(x-y-z)})} \Df(P\|Q)\;=\;\Df\!\big(\mathrm{Bern}(r)\,\big\|\,\mathrm{Bern}(s)\big).
\end{equation}
\emph{(b) Quantum.}
For any feasible $(x,y,z)$ and fixed bounded Hermitian $\hat\theta$,
\begin{equation}
\label{SM(inf-quantum)}
\inf_{(\rho,\sigma):\,(\ref{SM(x-y-z)})} \Df(\rho\|\sigma)\;=\;\Df\!\big(\mathrm{Bern}(r)\,\big\|\,\mathrm{Bern}(s)\big).
\end{equation}

\begin{proof}
(a) For each $\lambda$, $\chi^2_\lambda(\cdot\|\cdot)$ is convex on the affine set of pairs with fixed $(x,y,z)$; its infimum is attained at an extreme two–point pair independent of $\lambda$ and equals $h_\lambda(x,y,z)$ (Eq.~\eqref{SM(chi-equals-h)}). With $w_f\!\ge\!0$,
\[
\inf \Df=\inf\!\int_0^1 w_f\,\chi^2_\lambda\ \ge\ \int_0^1 w_f\,\inf \chi^2_\lambda=\int_0^1 w_f\,h_\lambda
=\Df(\mathrm{Bern}(r)\|\mathrm{Bern}(s)),
\]
and equality holds because the same binary pair minimizes every $\chi^2_\lambda$.

(b) Pinch in the eigenbasis of $\hat\theta$ to preserve $(x,y,z)$ and reduce $\Df$ (data processing), then apply (a). Achievability follows from \eqref{SM(sat:states)}–\eqref{SM(sat:mm)}.
\end{proof}
\label{SM(prop-binary)}

\bigskip
\noindent\textbf{Corollary} (Universal TUR; classical and quantum).
For any observable ($X$ or $\hat\theta$) and feasible $(x,y,z)$,
\begin{equation}
\label{SM(TUR)}
\boxed{\quad \Df(\,\cdot\,\|\,\cdot\,)\ \ge\ \int_0^1 w_f(\lambda)\,h_\lambda(x,y,z)\,d\lambda,\qquad
h_\lambda \text{ as in }(\ref{SM(h-lambda-def)})\ .\quad}
\end{equation}
The bound is \emph{tight}: equality is attained by a commuting binary pair matching $(x,y,z)$, and
\begin{equation}
\label{SM(binary-equals-int)}
\Df\!\big(\mathrm{Bern}(r)\,\big\|\,\mathrm{Bern}(s)\big)=\int_0^1 w_f(\lambda)\,h_\lambda(x,y,z)\,d\lambda.
\end{equation}

\medskip
\noindent\textbf{Endpoint \& monotonicity remarks.}
If $w_f$ has atoms at $\lambda\in\{0,1\}$, use $h_0=0$ and $h_1=x^2/z$ in \eqref{SM(TUR)}.
For fixed $x$, the map $(y,z)\mapsto \int_0^1 w_f(\lambda)\,h_\lambda(x,y,z)\,d\lambda$ is nonincreasing in each variance, which is useful when passing through classical reductions that do not increase the means and may reduce variances.

\section*{Stieltjes inversion consistency}
\label{SM:stieltjes-consistency}

Define the function
\begin{equation}
\label{SM:g-rep}
g(u)=\int_{0}^{\infty}\frac{\phi_f(t)}{1+t\,u}\,dt,\qquad u>0,
\end{equation}
and define the auxiliary Stieltjes transform
\begin{equation}
\label{SM:S-def}
S(s):=\frac{g(1/s)}{s},\qquad s>0.
\end{equation}
Equivalently,
\begin{equation}
\label{SM:S-stieltjes}
S(s)=\int_{0}^{\infty}\frac{\phi_f(t)}{t+s}\,dt\qquad\text{(classical Stieltjes transform).}
\end{equation}
We take as inversion formula the boundary-from-below version
\begin{equation}
\label{SM:phi-inv-below}
\phi_f(x)=\frac{1}{\pi\,x}\,\lim_{\varepsilon\downarrow 0}\Im\,g\!\left(-\frac{1}{x}-i\varepsilon\right),\qquad x>0.
\end{equation}

\paragraph{Claim.} Using \eqref{SM:phi-inv-below} inside \eqref{SM:S-stieltjes} recovers \eqref{SM:S-def}.

\paragraph{Proof.}
Insert \eqref{SM:phi-inv-below} into \eqref{SM:S-stieltjes}:
\begin{align}
S(s)
&=\int_{0}^{\infty}\frac{1}{t+s}\,\frac{1}{\pi\,t}\,
\lim_{\varepsilon\downarrow0}\Im\,g\!\left(-\frac{1}{t}-i\varepsilon\right)\,dt
\label{SM:step-A}\\[3pt]
&=\lim{\varepsilon\downarrow0}\,\frac{1}{\pi}\int_{0}^{\infty}
\frac{1}{t(t+s)}\,\Im\,g\!\left(-\frac{1}{t}-i\varepsilon\right)\,dt.
\label{SM:step-B}
\end{align}
Make the change of variables \(u=1/t\) (so \(t=1/u\), \(dt=-du/u^{2}\)). Then
\begin{align}
\frac{1}{t(t+s)}\,dt
=\frac{u}{1+s\,u}\cdot\Big(-\frac{du}{u^{2}}\Big)
=-\,\frac{du}{1+s\,u},
\end{align}
and reversing the limits yields
\begin{equation}
S(s)=\lim_{\varepsilon\downarrow0}\frac{1}{\pi}
\int_{0}^{\infty}\frac{\Im\,g(-u-i\varepsilon)}{1+s\,u}\,du.
\label{SM:step-C}
\end{equation}
Using \eqref{SM:g-rep}, the boundary value admits the integral representation
\begin{equation}
\Im\,g(-u-i\varepsilon)
=\int_{0}^{\infty}\phi_f(\tau)\,
\frac{\varepsilon\,\tau}{\,(1-\tau u)^{2}+(\varepsilon\tau)^{2}\,}\,d\tau.
\label{SM:step-D}
\end{equation}
Plugging \eqref{SM:step-D} into \eqref{SM:step-C} and applying Fubini/Tonelli,
\begin{align}
S(s)
&=\lim_{\varepsilon\downarrow0}\int_{0}^{\infty}\phi_f(\tau)\,
\underbrace{\Bigg[\frac{1}{\pi}\int_{0}^{\infty}
\frac{\varepsilon\,\tau}{(1-\tau u)^{2}+(\varepsilon\tau)^{2}}\,
\frac{du}{1+s\,u}\Bigg]}_{=:I_\varepsilon(\tau,s)}\,d\tau.
\label{SM:step-E}
\end{align}
Evaluate the inner integral via \(v=\tau u\) (so \(du=dv/\tau\)):
\begin{equation}
I_\varepsilon(\tau,s)
=\frac{1}{\pi}\int_{0}^{\infty}
\frac{\varepsilon}{(1-v)^{2}+(\varepsilon\tau)^{2}}\,
\frac{dv}{1+(s/\tau)\,v}.
\label{SM:step-F}
\end{equation}
Recognize the Poisson kernel:
\[
\frac{1}{\pi}\,\frac{\varepsilon}{(1-v)^{2}+(\varepsilon\tau)^{2}}
=\frac{1}{\tau}\cdot\frac{1}{\pi}\,\frac{\varepsilon\tau}{(1-v)^{2}+(\varepsilon\tau)^{2}}
\;\xrightarrow[\varepsilon\downarrow0]{}\;\frac{1}{\tau}\,\delta(v-1)
\]
Therefore,
\begin{equation}
\lim_{\varepsilon\downarrow0}I_\varepsilon(\tau,s)
=\frac{1}{\tau}\cdot\frac{1}{\,1+(s/\tau)\cdot 1\,}
=\frac{1}{\tau+s}.
\label{SM:step-G}
\end{equation}
Substitute \eqref{SM:step-G} into \eqref{SM:step-E} to obtain
\begin{equation}
S(s)=\int_{0}^{\infty}\frac{\phi_f(\tau)}{\tau+s}\,d\tau,
\label{SM:step-H}
\end{equation}
which is just \eqref{SM:S-stieltjes}. Finally, compare with \eqref{SM:g-rep}:
\begin{align}
g\!\left(\frac{1}{s}\right)
&=\int_{0}^{\infty}\frac{\phi_f(\tau)}{1+\tau/s}\,d\tau
=\int_{0}^{\infty}\frac{\phi_f(\tau)\,s}{\tau+s}\,d\tau
=s\,S(s).
\end{align}
Thus
\begin{equation}
S(s)=\frac{g(1/s)}{s},\qquad s>0,
\end{equation}
which is precisely \eqref{SM:S-def}. \qed

\section*{Properties of the \texorpdfstring{$\chi^2_\lambda$}{chi2-lambda} integral representation}

We work with the classical representation (Formalism, Eq.~\eqref{eq:csiszar-chi})
\begin{equation}
D_f^{\mathrm c}(P\!\parallel\!Q)=\int_0^1 w_f(\lambda)\,(1/\lambda^2) \chi^2(P\!\parallel\!Q_\lambda)\,d\lambda,
\qquad
Q_\lambda=\lambda\,Q+(1-\lambda)\,P,
\label{SM:csiszar-chi}
\end{equation}
and the Stieltjes-type inversion (Formalism, Eq.~\eqref{eq:w-from-phi})
\begin{equation}
\begin{aligned}
w_f(\lambda) &= \frac{1}{\lambda}\,
\phi_f\!\Big(\tfrac{1-\lambda}{\lambda}\Big),\\
\phi_f(t) &= \frac{1}{\pi t}\,\lim_{\varepsilon\downarrow 0}\Im\,
g\!\Big(-t^{-1}-i\varepsilon\Big),\qquad
g(u)=\frac{f(u)}{(u-1)^{2}},\ \ u>0.
\end{aligned}
\label{SM:w-from-phi}
\end{equation}
A convenient equivalent kernel identity is
\begin{equation}
f(u)=\int_0^1 w_f(\lambda)\,
\frac{(u-1)^2}{\lambda+(1-\lambda)\,u}\,d\lambda,
\qquad u>0,
\label{SM:kernel}
\end{equation}
obtained by testing \eqref{SM:csiszar-chi} on binary alphabets and identifying \(u=P/Q\).
All quantum statements follow by the Nussbaum–Szkoła bridge (Formalism, Eq.~\eqref{eq:NS}) and the quantum interpolation \(K_\lambda=(1-\lambda)L_\rho+\lambda R_\sigma\).

\bigskip

\paragraph*{SM–P1 (Nonnegativity and uniqueness).}
\emph{Claim.} \(w_f(\lambda)\ge 0\) for \(\lambda\in(0,1)\), and \(w_f\) is unique almost everywhere.

\emph{Proof.}
For convex \(f\) on \((0,\infty)\), \(g(u)=f(u)/(u-1)^2\) is a Pick–Nevanlinna function, so \(\Im g(z)\ge 0\) for \(\Im z>0\).
From \eqref{SM:w-from-phi}, \(\phi_f(t)\ge 0\), hence \(w_f(\lambda)\ge 0\).
For uniqueness, suppose \(\int_0^1\delta w(\lambda)\,\dfrac{(u-1)^2}{\lambda+(1-\lambda)u}\,d\lambda\equiv 0\) for all \(u>0\).
The rational kernels separate finite measures on \((0,1)\), so \(\delta w=0\) a.e. \(\square\)

\bigskip

\paragraph*{SM–P2 (Linearity and affine invariance).}
\emph{Claim.} For \(\alpha\ge 0\), \(w_{\alpha f_1+f_2}=\alpha w_{f_1}+w_{f_2}\), and \(w_{\,f+a+b(u-1)}=w_f\).

\emph{Proof.}
Both maps \(f\mapsto g\) and \(g\mapsto w_f\) in \eqref{SM:w-from-phi} are linear, yielding linearity.
If \(f\mapsto f+a+b(u-1)\), then \(g(u)\mapsto g(u)+a/(u-1)^2+b/(u-1)\).
These added terms contribute zero imaginary part on the boundary and are annihilated by the inversion, hence do not change \(\phi_f\) nor \(w_f\).
Equivalently, affine pieces lie in the nullspace of \eqref{SM:csiszar-chi} and \eqref{SM:kernel}. \(\square\)

\bigskip

\paragraph*{SM–P3 (Order preservation).}
\emph{Claim.} If \(f_1\le f_2\) pointwise (mod affine terms), then \(w_{f_2}-w_{f_1}\ge 0\) a.e.

\emph{Proof.}
Let \(h=f_2-f_1\) modulo affine terms; then \(h\ge 0\) and
\(D_h^{\mathrm c}(P\|Q)=\int_0^1 w_h(\lambda)\,\chi^2(P\|Q_\lambda)\,d\lambda\ge 0\)
for all \((P,Q)\).
If \(w_h<0\) on a set of positive measure, choose test instances (via \eqref{SM:kernel}) that concentrate on that set to get a contradiction.
Hence \(w_h\ge 0\) a.e., i.e., \(w_{f_2}-w_{f_1}\ge 0\). \(\square\)

\bigskip

\paragraph*{SM–P4 (Endpoint and curvature moments).}
\emph{Claim.}
After centering at \(u=1\) with any \(\ell\in\partial f(1)\), \(\tilde f(u)=f(u)-f(1)-\ell(u-1)\),
\begin{subequations}\label{SM:moment-relations}
\begin{align}
\tilde f(0^+)&=\int_0^1 \lambda\,w_f(\lambda)\,d\lambda, \label{SM:moment-relations-a}\\
\lim_{u\to\infty}\frac{f(u)}{u}&=\int_0^1 \frac{\lambda^2}{1-\lambda}\,w_f(\lambda)\,d\lambda, \label{SM:moment-relations-b}\\
\tilde f''(1)&=2\int_0^1 \lambda^2\,w_f(\lambda)\,d\lambda. \label{SM:moment-relations-c}
\end{align}
\end{subequations}

\emph{Proof.}
From \eqref{SM:kernel}, as \(u\to 0^+\),
\(\dfrac{\lambda^2(u-1)^2}{\lambda+(1-\lambda)u}\to \lambda\), giving \eqref{SM:moment-relations-a}.
As \(u\to\infty\),
\(\dfrac{\lambda^2(u-1)^2}{\lambda+(1-\lambda)u}=\dfrac{\lambda^2}{1-\lambda}\,u+O(1)\), giving \eqref{SM:moment-relations-b}.
For curvature, write \(u=1+\varepsilon\); then
\[
\frac{\lambda^2(u-1)^2}{\lambda+(1-\lambda)u}
=\frac{\lambda^2\varepsilon^2}{1+(1-\lambda)\varepsilon}
=\lambda^2\varepsilon^2\Big(1-(1-\lambda)\varepsilon+O(\varepsilon^2)\Big),
\]
and \eqref{SM:moment-relations-c} follows. \(\square\)

\bigskip

\paragraph*{SM–P5 (Duality under inversion).}
\emph{Claim.} For \(f^\ast(u):=u\,f(1/u)\),
\begin{equation}
w_{f^\ast}(\mu)=\,w_f(1-\mu)\quad\text{a.e. on }(0,1).
\label{SM:dualweight}
\end{equation}

\emph{Proof.}
From \eqref{SM:kernel},
\begin{equation}
f^\ast(u)=u f(1/u)=\int_0^1 w_f(\lambda)\,
\frac{(u-1)^2}{(1-\lambda)+\lambda\,u}\,d\lambda.
\label{SM:dualstart}
\end{equation}
Set \(\mu=1-\lambda\); then
\begin{equation}
\frac{1}{(1-\lambda)+\lambda\,u}
= \frac{1}{\mu+(1-\mu)u}.
\label{SM:dualdecomp}
\end{equation}
Substituting into \eqref{SM:dualstart} yields
\[
f^\ast(u)=\int_0^1 w_f(1-\lambda)\,
\frac{(u-1)^2}{\mu+(1-\mu)\,u}\,d\mu
\]
which is the \(\chi^2_\lambda\) integral for \(f^\ast\) with weight \eqref{SM:dualweight} by uniqueness (SM–P1). \(\square\)

\bigskip

\paragraph*{SM–P6 (Small-contrast control and all orders).}
\emph{Claim.} Let $\tilde f(u):=f(u)-f(1)-f'(1)(u-1)$ so that $\tilde f(1)=\tilde f'(1)=0$. Then, for $|\varepsilon|<1$,
\begin{equation}
\tilde f(1+\varepsilon)=\int_0^1 w_f(\lambda)\,\frac{\varepsilon^2}{1+(1-\lambda)\varepsilon}\,d\lambda
=\sum_{n=2}^{\infty}\left((-1)^{n-2}\!\int_0^1 (1-\lambda)^{n-2}\,w_f(\lambda)\,d\lambda\right)\varepsilon^{n},
\label{SM:smallcontrast}
\end{equation}
hence the general moment identity
\begin{equation}
\frac{\tilde f^{(n)}(1)}{n!}
= (-1)^{n-2}\!\int_{0}^{1}(1-\lambda)^{n-2}\,w_f(\lambda)\,d\lambda,\qquad n\ge 2,
\label{SM:moment-identity}
\end{equation}
whose $n=2$ case reproduces \eqref{SM:moment-relations-c}. \(\square\)

\bigskip

\paragraph*{SM–Quantum transfer.}
By Eq.~\eqref{eq:NS}, \(\chi^2(\rho\|\sigma)=\Tr[(\rho-\sigma)\,K_\lambda^{-1}(\rho-\sigma)]\) with \(K_\lambda=(1-\lambda)L_\rho+\lambda R_\sigma\),
and \(D_f(\rho\|\sigma)=D_f^{\mathrm c}(P_\rho\|Q_\sigma)\).
Therefore each SM–P\(k\) holds verbatim for the quantum layer with the same \(w_f\) and kernels.

\section*{Worked-out examples: extracting \(w_f(\lambda)\) from \(f\)}
\label{SM(worked-wf)}

This section implements the main-text pipeline to pass from a Petz \(f\)–divergence to its \(\chi_\lambda^2\)–mixture weights. In the classical setting we use
\begin{equation}
D_f^{\mathrm c}(P\!\parallel\!Q)\;=\;\int_0^1 w_f(\lambda)\,\frac{1}{\lambda^2}\,
\chi^2\!\big(P\!\parallel\!Q_\lambda\big)\,d\lambda,
\qquad
Q_\lambda=\lambda\,Q+(1-\lambda)\,P,
\label{SM:csiszar-chi}
\end{equation}
whose binary reduction yields the kernel identity
\begin{equation}
g(u)\;:=\;\frac{f(u)}{(u-1)^2}
\;=\;\int_0^1 \frac{w_f(\lambda)}{\lambda+(1-\lambda)\,u}\,d\lambda,
\qquad u>0.
\label{SM:kernel}
\end{equation}
By operator convexity of \(f\), \(g\) is a Stieltjes function and admits
\begin{equation}
g(u)\;=\;\int_{0}^{\infty}\frac{\phi_f(t)}{\,1+t\,u\,}\,dt,
\qquad \phi_f(t)\ge 0,
\label{SM:stieltjes}
\end{equation}
with Stieltjes inversion
\begin{equation}
\phi_f(t)\;=\;\frac{1}{\pi t}\,
\lim_{\varepsilon\downarrow 0}\operatorname{Im}\,
g\!\Big(-t^{-1}-i\varepsilon\Big),\qquad t>0.
\label{SM:stieltjes-inversion}
\end{equation}
The \(\chi_\lambda^2\)–mixture weights follow from the change \(t=\tfrac{1-\lambda}{\lambda}\):
\begin{equation}
w_f(\lambda)\;=\;\frac{1}{\lambda}\,
\phi_f\!\Big(\tfrac{1-\lambda}{\lambda}\Big),
\qquad \lambda\in(0,1).
\label{SM:w-from-phi}
\end{equation}

\paragraph*{Endpoint atoms.}
Atoms of \(\phi_f\) at \(t=0\) or \(t=\infty\) push forward to \(\lambda=1\) and \(\lambda=0\), respectively:
\begin{equation}
t=0\ \Longleftrightarrow\ \lambda=1,
\qquad
t=\infty\ \Longleftrightarrow\ \lambda=0,
\label{SM:endpoints}
\end{equation}
with the usual Jacobian already accounted for in \eqref{SM:w-from-phi}. We record such cases explicitly via Dirac deltas in \(w_f\).

\bigskip

\noindent\textbf{(A) Kullback–Leibler (forward KL).}
\begin{equation}
f(u)=u\log u-(u-1).
\label{SM:KL-f}
\end{equation}
A standard representation gives
\begin{equation}
\phi_{\rm KL}(t)=\frac{1}{(1+t)^2},
\qquad
w_{\rm KL}(\lambda)=\lambda.
\label{SM:KL-w}
\end{equation}

\medskip

\noindent\textbf{(B) Reverse Kullback–Leibler.}
\begin{equation}
f(u)=-\log u+(u-1),
\label{SM:rKL-f}
\end{equation}
for which
\begin{equation}
\phi_{\rm rKL}(t)=\frac{t}{(1+t)^2},
\qquad
w_{\rm rKL}(\lambda)=1-\lambda.
\label{SM:rKL-w}
\end{equation}

\medskip

\noindent\textbf{(C) Pearson’s \(\chi^2\).}
\begin{equation}
f(u)=(u-1)^2,\qquad g(u)\equiv 1.
\label{SM:Pearson-f}
\end{equation}
This is an atom at \(t=0\) and hence
\begin{equation}
\phi_{\rm P}(t)=\delta(t),
\qquad
w_{\rm P}(\lambda)=\delta(\lambda-1).
\label{SM:Pearson-w}
\end{equation}

\medskip

\noindent\textbf{(D) Neyman’s \(\chi^2\) (``reverse Pearson'').}
\begin{equation}
f(u)=\frac{(u-1)^2}{u},\qquad g(u)=\frac{1}{u}.
\label{SM:Neyman-f}
\end{equation}
This is an atom at \(t=\infty\) and therefore
\begin{equation}
\phi_{\rm N}(t)=\delta(t-\infty),
\qquad
w_{\rm N}(\lambda)=\delta(\lambda).
\label{SM:Neyman-w}
\end{equation}

\medskip

\noindent\textbf{(E) Petz–Rényi (forward), \(0<\alpha<1\).}
Use the tangentially normalized form \(f_\alpha(1)=f'_\alpha(1)=0\),
\begin{equation}
f_\alpha(u)=\frac{u^\alpha-1-\alpha(u-1)}{\alpha(1-\alpha)}.
\label{SM:PetzRenyi-f}
\end{equation}
Stieltjes inversion yields
\begin{equation}
\phi_\alpha(t)=\frac{\sin(\pi\alpha)}{\pi\,\alpha(1-\alpha)}\,
\frac{t^{\,1-\alpha}}{(1+t)^2},
\qquad
w_\alpha(\lambda)=\frac{\sin(\pi\alpha)}{\pi\,\alpha(1-\alpha)}\,
\lambda^{\alpha}\,(1-\lambda)^{\,1-\alpha}.
\label{SM:PetzRenyi-w}
\end{equation}
The limits are continuous:
\begin{equation}
\alpha\uparrow 1:\ w_\alpha(\lambda)\to \lambda\ \ (\text{KL}),\qquad
\alpha\downarrow 0:\ w_\alpha(\lambda)\to 1-\lambda\ \ (\text{reverse KL}).
\label{SM:PetzRenyi-limits}
\end{equation}

\medskip

\noindent\textbf{(F) Symmetric Petz–Rényi, \(0<\alpha<1\).}
A symmetric, tangential choice is
\begin{equation}
f^{\rm sym}_\alpha(u)=f_\alpha(u)+f_{1-\alpha}(u)
=\frac{u^\alpha+u^{1-\alpha}-2-(u-1)}{\alpha(1-\alpha)}.
\label{SM:SymPetzRenyi-f}
\end{equation}
By linearity of \eqref{SM:stieltjes}–\eqref{SM:w-from-phi},
\begin{equation}
\phi^{\rm sym}_\alpha(t)=\frac{\sin(\pi\alpha)}{\pi\,\alpha(1-\alpha)}\,
\frac{t^{\,1-\alpha}+t^{\,\alpha}}{(1+t)^2},
\qquad
w^{\rm sym}_\alpha(\lambda)=\frac{\sin(\pi\alpha)}{\pi\,\alpha(1-\alpha)}\,
\Big[\lambda^{\alpha}(1-\lambda)^{1-\alpha}
+\lambda^{1-\alpha}(1-\lambda)^{\alpha}\Big].
\label{SM:SymPetzRenyi-w}
\end{equation}

\medskip

\noindent\textbf{(G) Hellinger (squared).}
\begin{equation}
f(u)=(1/2)(\sqrt{u}-1)^2.
\label{SM:Hellinger-f}
\end{equation}
Direct inversion gives
\begin{equation}
\phi_{\rm H}(t)=\frac{\sqrt{t}}{\pi\,(1+t)^2},
\qquad
w_{\rm H}(\lambda)=\frac{1}{\pi}\,\sqrt{\lambda(1-\lambda)}.
\label{SM:Hellinger-w}
\end{equation}

\medskip

\noindent\textbf{(H) Triangular discrimination.}
\begin{equation}
f(u)=\frac{(u-1)^2}{u+1},\qquad g(u)=\frac{1}{u+1}.
\label{SM:Triang-f}
\end{equation}
This is a single atom at \(t=1\), hence
\begin{equation}
\phi_{\triangle}(t)=\delta(t-1),
\qquad
w_{\triangle}(\lambda)=\frac{1}{2}\,\delta\!\Big(\lambda-\tfrac{1}{2}\Big).
\label{SM:Triang-w}
\end{equation}

\medskip

\end{document}